\documentclass{article}
\usepackage{spconf,amsmath,graphicx}
\usepackage{makecell}
\usepackage{amsfonts}

\usepackage{flushend}
\usepackage{algorithm}
\usepackage{algorithmicx}
\usepackage{mathtools}
\usepackage{algpseudocode}
\usepackage{subfigure}
\usepackage{multirow}
\usepackage{longtable}
\usepackage{amsthm}
\usepackage{amssymb }

\usepackage{bm}
\usepackage{mathrsfs}

\usepackage{longtable}
\usepackage{multirow}
\usepackage{cite}
\usepackage{xcolor}
\usepackage{graphicx}

\def\M{M}
\def\SM{\text{CAET-SWin}}
\def\Q{\bm{Q}}
\def\K{\bm{K}}
\def\V{\bm{V}}
\def\W{\bm{W}}
\title{Spatio-Temporal Hybrid Fusion of CAE and SWIn Transformers for Lung Cancer Malignancy Prediction}
\name{
\makecell{Sadaf Khademi$^\dagger$, Shahin Heidarian$^\ddagger$,
Parnian Afshar$^\dagger$, Farnoosh Naderkhani$^\dagger$
Anastasia Oikonomou$^{\dagger\dagger}$,\\
Konstantinos N. Plataniotis$^{\ddagger\ddagger}$,
and Arash Mohammadi$^\dagger$}}
\address{$~^\dagger$Concordia Institute for Information Systems Engineering, Concordia University, Montreal, Canada\\
$~^\ddagger$Department of Electrical and Computer Engineering, Concordia University, Montreal, Canada \\
$~^{\dagger\dagger}$Department of Medical Imaging, Sunnybrook Health Sciences Centre, Toronto, Canada\\
$~^{\ddagger\ddagger}$Department of Electrical and Computer Engineering, University of Toronto, Toronto, Canada}
\begin{document}
\ninept
\maketitle
%
\begin{abstract}
The paper proposes a novel hybrid discovery Radiomics framework that simultaneously integrates temporal and spatial features extracted from non-thin chest Computed Tomography (CT)  slices to predict Lung Adenocarcinoma (LUAC) malignancy with minimum expert involvement.
Lung cancer is the leading cause of mortality from cancer worldwide and has various histologic types, among which LUAC has recently been the most prevalent. LUACs are classified as pre-invasive, minimally invasive, and invasive adenocarcinomas. Timely and accurate knowledge of the lung nodules malignancy leads to a proper treatment plan and reduces the risk of unnecessary or late surgeries. Currently, chest CT scan is the primary imaging modality to assess and predict the invasiveness of LUACs. However, the radiologists' analysis based on CT images is subjective and suffers from a low accuracy compared to the ground truth pathological reviews provided after surgical resections.
The proposed hybrid framework, referred to as the $\SM$, consists of two parallel paths: (i) The Convolutional Auto-Encoder (CAE) Transformer path that extracts and captures informative features related to inter-slice relations via a modified Transformer architecture, and; (ii) The Shifted Window (SWin) Transformer path, which is a hierarchical vision transformer that extracts nodules' related spatial features from a volumetric CT scan. Extracted temporal (from the CAET-path) and spatial (from the Swin path) are then fused through a fusion path
to classify LUACs.
Experimental results on our in-house dataset of $114$ pathologically proven Sub-Solid Nodules (SSNs) demonstrate that the $\SM$ significantly improves reliability of the invasiveness prediction task while achieving an accuracy of $82.65$\%, sensitivity of $83.66$\%, and specificity of $81.66$\% using 10-fold cross-validation.
\end{abstract}
%
%
\vspace{-.15in}
\section{Introduction} \label{sec:intro}
\vspace{-.1in}
Lung Cancer (LC) is the deadliest and least-funded cancer worldwide~\cite{Kamath2019,Bray2018}. Non-small-cell LC is the major type of LC, and Lung Adenocarcinoma (LUAC) is the most prevalent histologic sub-type~\cite{Herbst2018}. Lung nodules manifesting as Ground Glass (GG) or Subsolid Nodules (SSNs) on Computed Tomography (CT) scans have a higher risk of malignancy than other incidentally detected small solid nodules. SSNs are often diagnosed as adenocarcinoma and are generally classified into pure GG nodules and part-solid nodules according to their appearance on the lung window settings~\cite{Kim2007, Lai2021}. A timely and accurate attempt to differentiate the LUACs is of utmost importance to guide a proper treatment plan, as in some cases, a pre-invasive or minimally invasive SSN can be monitored with regular follow-up CT scans, whereas invasive lesions should undergo immediate surgical resection if they are deemed eligible. Most often, the SSNs type is diagnosed based on the pathological findings performed after surgical resections, which is not desired for prior treatment planning. Currently, radiologists use chest CT scans to assess the invasiveness of the SSNs based on their imaging findings and patterns prior to making decisions regarding the appropriate treatment. Such visual approaches, however, are time-consuming, subjective, and error-prone. So far, many studies have used high-resolution and thin-slice ($<1.5mm$) CT images for the SSN classification, which require longer analysis times, as well as more reconstruction time~\cite{Cui2020, Shao2020}. However, lung nodules are mostly identified from CT scans performed for varied clinical purposes acquired using routine standard or low-dose scanning protocols with non-thin slice thicknesses (up to $5mm$)~\cite{Oikonomou2019}. In addition, recent lung cancer screening recommendation, suggests using low-dose CT scans with thicker slice-thicknesses (up to $2.5mm$)~\cite{Kazerooni2014,Fujii2017}. Capitalizing on the above discussion, the necessity of developing an automated invasiveness assessment framework that performs well regardless of technical settings has recently arisen among the research community and healthcare professionals.

\noindent
\textbf{Related Works:} Generally speaking, existing works on the SSN invasiveness assessment can be categorized into two main classes: (i) Radiomics-based, and; (ii) Deep Learning-based frameworks, also referred to as Discovery Radiomics~\cite{Gu2021}. In the former class, data-characterization algorithms extract quantitative features from nodule masks and the original CT images, which are then analyzed using statistical or conventional Machine Learning (ML) models~\cite{Gao2019,Uthoff2019}. As an example of such frameworks, a histogram-based model is developed in~\cite{Oikonomou2019} to predict the invasiveness of primary adenocarcinoma SSNs from non-thin CT scans of $109$ pathologically labeled SSNs. In this study, a set of histogram-based and morphological features along with additional features extracted via the functional Principal Component Analysis (PCA) is fed to a linear logistic regression.
Discovery Radiomics approaches, on the other hand, extract informative and discriminative features in an automated fashion. Existing deep models working with volumetric CT scans can be classified in two categories: (i) 3D-based solutions~\cite{Liu2017}, where the whole 3D volume of CT images are fed to the model. Processing a large 3D CT scan at once, however, results in extensive computational complexity requiring more computational resources, and enormous training datasets, and; (ii) 2D-based solutions~\cite{Heidarian2021,Heidarian2021a,Farhangi2020}, where individual 2D CT slices are first analyzed, which are then fused via an aggregation mechanism to represent characteristics of the whole volume.

Due to the time-series nature of the volumetric CT scans, which utilize a sequence of 2D images to provide a detailed representation of the organ, there has been recently a surge of interest in Category (ii), especially, in the application of sequential deep models for diagnostic/prognostic tasks based on 2D-CT scans.
In 2017, a new sequential deep model, commonly known as ``Transformer"~\cite{Vaswani2017}, has been proposed illustrating superior performance in the tasks with sequential input data. Transformer models benefit from a novel self-attention mechanism, which is capable of capturing global context and dependencies between instances of the sequential data while requiring far less computational resources compared to conventional recurrent-based architectures such as Recurrent Neural Network (RNN) and Long Short-Term Memory (LSTM). Transformers are also superior to their counterparts in terms of parallelization and dynamic attention. Although the transformer model was initially designed for natural language processing, in 2020 Vision Transformer (ViT) was introduced to adopt the self-attention mechanism for image processing applications~\cite{Dosovitskiy2020}. ViT-based models apply the self-attention mechanism to the sequence of patches extracted from a 2D image. It is worth mentioning that development of transformer -based models in processing sequential medical images is progressing rapidly especially models~\cite{Gao2021,Ambita2021} for COVID-19  identification and segmentation have shown promising results and potentials.

\vspace{.025in}
\noindent
\textbf{Contributions:}
The paper proposes a novel hybrid malignancy predictive framework, referred to as the $\SM$, which combines spatial and temporal features extracted by two parallel self-attention mechanisms (the CAET and SWin paths). Intuitively speaking, the $\SM$ is designed to take advantage of the 3D nature of non-thin CT sequences by jointly modeling temporal (inter-slice) and spatial (within-slice) variations of CT slices with reduced computational complexity.
Each of the constituent paths concentrates on a specific domain to capture local and global evidence of invasive nodules. The first path is built based on a Convolutional Auto-Encoder (CAE) algorithm~\cite{Masci2011} to form a sequential feature map of the whole lung provided in a set of CT slices. The obtained sequential feature maps are then directly fed to the encoder part of a transformer model containing multiple Multi-head Self Attention (MSA) layers to find temporal dependencies between slices. This path of the $\SM$  has an intuitively pleasing modified structure, i.e., the patch flattening and linear projection components of the ViT architecture are removed and the output of the CAE is directly fed to the Position Embedding (PE) module.
At the same time, nodule patches with each slice are analyzed in the second parallel path, which is a ViT with a hierarchical structure using the shifted windowing scheme of the SWin transformer. This path finds spatial connections among local windows formed for each image that significantly enhances spatial modeling power~\cite{Liu_2021_ICCV}. Finally, temporal and spatial features extracted in these parallel paths are fused via a stack of fully connected (FC) layers to provide the final predictions.
An important aspect considered in the design of the proposed hybrid $\SM$ framework is that it does not require a detailed annotation of the nodules needed in most existing models, which is a challenging and time-consuming task even for expert radiologists. The only required information from the radiologists/experts is the set of slices with evidence of a nodule and an approximate region containing a suspicious nodule.

\vspace{-.175in}
\section{Materials and Methods} \label{sec:MM}
\vspace{-.1in}
\setlength{\textfloatsep}{0pt}
\begin{figure}[t!]
    \centering
    \includegraphics[width=0.85\columnwidth]{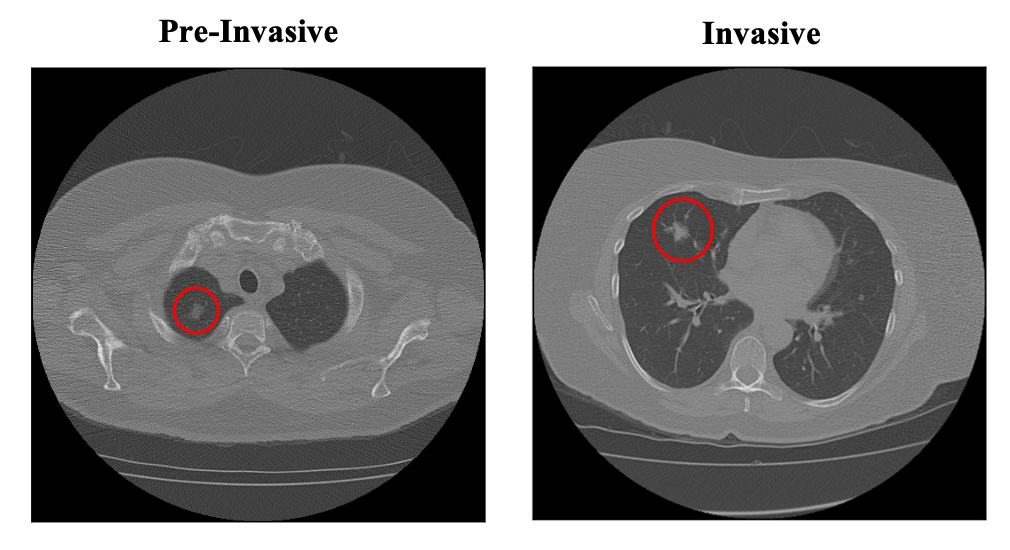}
\vspace{-.2in}
    \caption{\footnotesize Sample pre-invasive and invasive adenocarcinomas.}
    \label{fig:sample-ct}
\end{figure}
In this section, first, we briefly present the dataset utilized to develop and test the proposed $\SM$ framework. Afterwards, we describe the lung segmentation module used as a pre-processing step to form the required input of the two parallel paths of the $\SM$ framework. Finally, we briefly introduce the self-attention mechanism, which is the main building block of both parallel paths of our hybrid framework.

\vspace{.025in}
\noindent
\textbf{2.1. Dataset}

\noindent
Most of recent studies~\cite{Parnian:SREP2020} were developed and evaluated based on the public LIDC-IDRI~\cite{Armato2011} dataset, which does not have pathologically proven labels and focuses more on nodule detection than classification. In this study, we have used the dataset initially introduced in~\cite{Oikonomou2019} and added five additional cases acquired from the same institution to further balance the dataset. This dataset contains non-thin volumetric CT scans of $114$ pathologically proven SSNs (with technical parameters of $100–135$ kVp and $80–120$ mAs) segmented and reviewed by $2$ experienced thoracic radiologists. All SSN labels are provided after surgical resections. SSNs are initially classified according to their histology into three categories of pre-invasive lesions including atypical adenomatous hyperplasia and adenocarcinoma in situ, minimally invasive, and invasive pulmonary adenocarcinoma~\cite{Lai2021}. Following the original study~\cite{Oikonomou2019}, we have grouped the first two categories to represent the pre-invasive and minimally invasive class with $58$ cases, and kept the invasive nodules as the other class including $56$ cases. Fig.~\ref{fig:sample-ct} shows two sample lung adenocarcinomas from the dataset.

\vspace{.025in}
\noindent
\textbf{2.2. Lung Segmentation}

\noindent
As the pre-processing step, we have utilized a well-trained U-Net-based lung segmentation model, introduced in~\cite{Hofmanninger2020}, to extract the lung parenchyma from the CT scans. This approach has been proven beneficial to enhance the learning process and final results of deep learning-based models in previous CT-related studies~\cite{Heidarian2021,Afshar2021,Mohammadi2021}, by removing distracting components from the CT images. The extracted lung areas are then passed into two parallel paths (i.e., the CAET and the SWin paths) of the $\SM$ framework as follows: (i) \textit{Input of the CAET Path:} Extracted lung areas after the pre-processing step are down-sampled from ($512, 512$) to ($256, 256$) pixels to reduce the complexity and memory allocation without significant loss of information. (ii) \textit{Input of the SWin Path:} Nodule patches within the segmented area resulting from the pre-processing step based on SSN annotation provided by radiologists and then zero-padded to ($224, 224$) pixels.

\vspace{.025in}
\noindent
\textbf{2.3. Multi-Head Self-Attention Mechanism}

\noindent
Transformer architecture constitutes the main component of the two parallel paths of the $\SM$ framework, which uses a self-attention mechanism to capture dependencies among various instances of the input sequence. Self-attention is developed based on a scaled dot-product attention function, mapping a query and a set of key-value pairs to an output. The query ($Q$), keys ($K$), and values ($V$) are learnable representative vectors for the instances in the input sequence with dimensions $d_k$, $d_k$, and $d_v$, respectively. The output of the self-attention module is computed as a weighted average of the values, where the weight assigned to each value is computed by a similarity function of the query and the corresponding key after applying a softmax function~\cite{Vaswani2017}. More specifically, the attention values on a set of queries are computed simultaneously, packed together into matrix $\Q$. The keys and values are similarly represented by matrices $\K$ and $\V$. The output of the attention function is computed~as follows
\vspace{-.125in}
\begin{equation}
    Attention(\Q,\K,\V) = \text{softmax}\left(\frac{\Q\K^{T}}{\sqrt{d_k}}\right)\V,\vspace{-.1in}
    \label{eq:attention}
\end{equation}
where superscript $^{T}$ denotes transpose of a given matrix. It is also beneficial to linearly project the queries, keys, and values $h$ times with various learnable linear projections to vectors with $d_k$, $d_k$ and $d_v$ dimensions, respectively, before applying the attention function. On each of the projected versions of queries, keys, and values, the attention function is performed in parallel, resulting in $d_{v}-dimensional$ output values. These values are then concatenated and once again linearly projected via a FC layer. This process is called Multi-head Self Attention (MSA), which helps the model to jointly attend to information from different representation subspaces at different positions~\cite{Vaswani2017}. The output of the MSA module is
\begin{eqnarray}
MSA(\Q,\K,\V) &\!\!\!\!=\!\!\!\!& Concat\big(head_1, \cdots, head_h\big) \W^O, \nonumber \\
\!\!\!\!\text{where }\quad\ head_i &\!\!\!\!=\!\!\!\!& Attention(\Q\W_i^Q,\K\W_i^K,\V\W_i^V),\label{eq:mha}
\end{eqnarray}
where the projections are achieved by parameter matrices $\W_i^Q \in \mathbb{R}^{d_{model}\times d_k}$, $\W_i^K \in \mathbb{R}^{d_{model}\times d_k}$, $\W_i^V \in \mathbb{R}^{d_{model}\times d_v}$, and $\W^O \in \mathbb{R}^{{hd_v}\times d_{model}}$.
This completes an overview of the preliminary material, next, we develop the $\SM$ framework.

\vspace{-.15in}
\section{The Proposed $\SM$ Framework}
\vspace{-.1in}
\begin{figure}[t!]
    \centering
    \includegraphics[scale=.25]{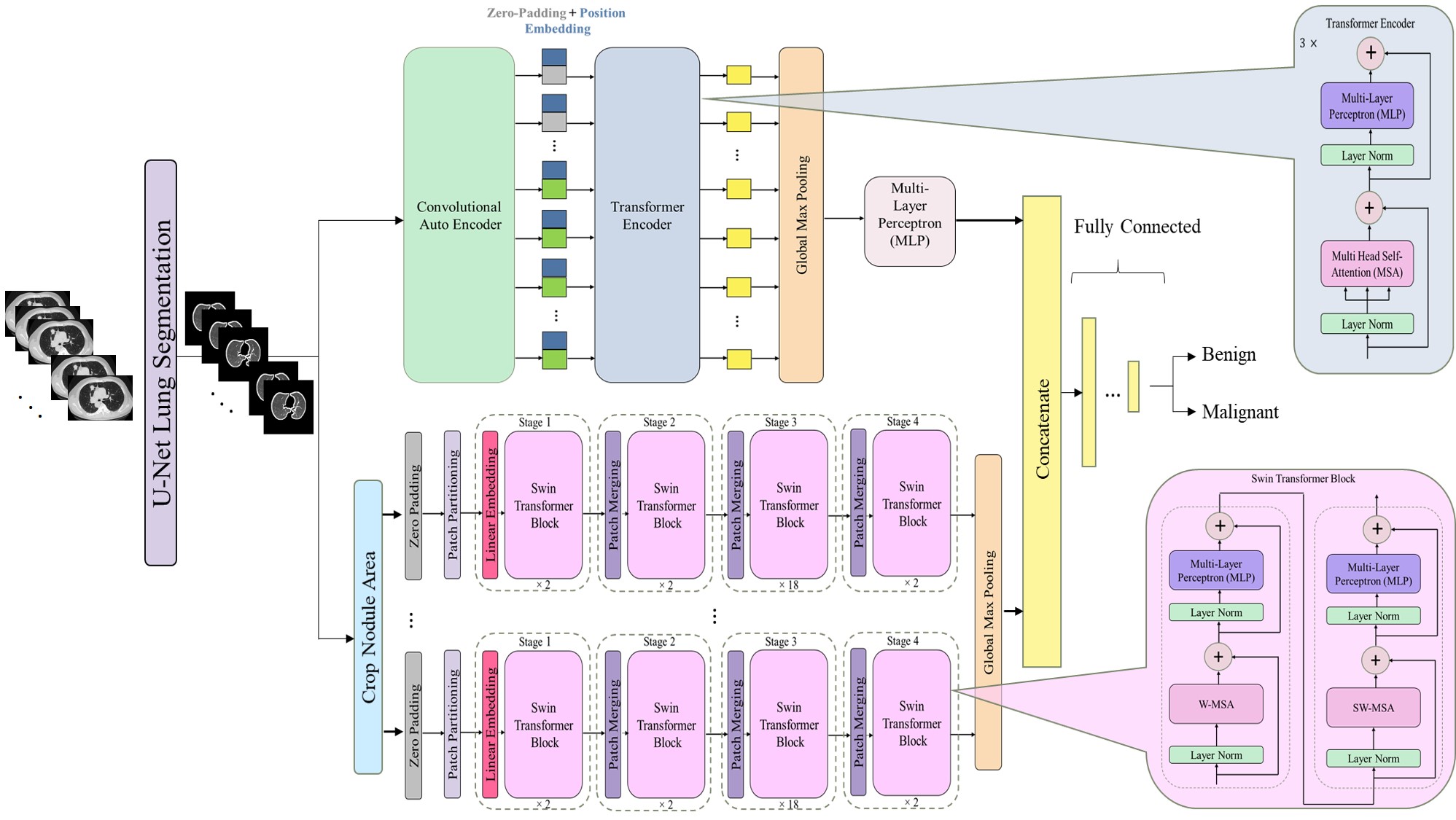}
    \vspace{-.15in}
    \caption{\footnotesize Pipeline of the $\SM$ Transformer.}
    \label{fig:cae-transformer}
\end{figure}
The proposed hybrid $\SM$ framework combines spatial and temporal features extracted by two parallel self-attention mechanisms (the so called CAET and Swin paths) to perform malignancy prediction based on CT images. Through its hybrid architecture, the $\SM$ learns from the 3D structure of non-thin CT scans by simultaneous extraction of inter-slice (through its CAET path) and within-slice (through its SWin path) features. Extracted features are then fused through a FC fusion path (implemented in series to the CAET and SWin parallel paths) to form the output. In what follows, we present each of the three paths of the $\SM$ framework.

\vspace{.05in}
\noindent
\textbf{3.1. The CAE-Transformer Path}

\noindent
To extract temporal (inter-slice) features, the first feature extractor module of the proposed hybrid $\SM$ is named CAE-Transformer, which integrates CAE and a modified version of the transformer encoder~\cite{Vaswani2017,Dosovitskiy2020}. Fig.~\ref{fig:cae-transformer} illustrates the pipeline of the CAE-Transformer framework, along with the architecture of a transformer encoder. To develop the CAE-Transformer module, we initially pre-trained a CAE model based on CT images with the evidence of a nodule in public LIDC-IDRI dataset. This is performed to represent CT images with compressed and informative feature maps. The CAE model consists of an encoder and a decoder part, where the former is responsible for generating a compact representation of the input image through a stack of five convolution and five max-pooling layers (kernel size of $2\times2$) followed by a FC layer with the size of $256$. The decoder component aims to reconstruct the original image using the compressed feature representation generated by the encoder. By minimizing the Mean Squared Error (MSE) between the original and the reconstructed image, the CAE learns to produce highly informative feature representations for the input images. The pre-trained model on the LIDC-IDRI dataset is then fine-tuned on our in-house dataset (presented in Section~\ref{sec:MM}).

The CAE component plays the role of the embedding layer in basic transformer architecture. In other words, instead of patch flattening and linear projection, which is commonly used in the transformer architecture, the output of the CAE is directly fed to the PE module. As the number of slices with the evidence of a nodule varies between different subjects (from $2$ to $25$ slices per nodule), we have taken the maximum number of slices in our dataset (i.e., $25$ slices) and zero-padded the input sequences based on this number such that all sequences provided by CAE would have the same dimension of ($25,256$). The PE layer is then incorporated into the model to add information about the position of slices in the input sequence. More specifically, a transformer encoder is initialized by applying the MSA on the normalized CAE-generated feature maps corresponding to the input instances, followed by a residual connection, which adds low-level features of the input to the output of the MSA module. A Layer Normalization (LN) is then applied to the results. The normalized values are then passed to the next module, which contains a Multi-Layer Perceptron (MLP), followed by another residual connection as shown in Fig.~\ref{fig:cae-transformer}.

The CAE-Transformer path is constructed by stacking $3$ transformer encoder blocks on top of each other with a projection dimension of $256$, key and query dimensions of $128$, and $5$ heads in each MSA module. Finally, the outputs obtained by stack of transformer encoders from all input instances (slices) are passed to a Global Max Pooling (GMP) layer followed by an MLP layer with $32$ neurons and a Relu activation function to generate the feature vectors.

\vspace{.025in}
\noindent
\textbf{3.2. The SWin-Transformer Path}
\begin{table}[t!]
    \centering
     \caption{\footnotesize Results obtained by the CAET-SWin and its counterparts.    \label{tab:results}}
     \resizebox{\columnwidth}{!}{
    \begin{tabular}{|c|c|c|c|}
    \hline
        \textbf{Model} & \textbf{Accuracy [95\%CI]} & \textbf{Sensitivity \%} & \textbf{Specificity \%}\\
    \hline
        Ref.~\cite{Oikonomou2019} & 81.00 [58.1 94.6] & 80.00 & \textbf{81.80}\\
    \hline
        Ref.~\cite{Gao2021} & 61.66 [49.9 73.5] & 61.33 & 62.33\\
    \hline
        CAET (GAP) & 56.21 [46.7 65.7] & 42.00 & 69.66\\
    \hline
        CAET (Flatten) & 64.84 [57.9 71.7] & 54.66 & 74.66\\
    \hline
        CAET (GMP) & 69.46 [58.8 80.2] & 64.33 & 74.66\\
    \hline
        SWin & 78.10 [70.0 86.2] & 76.66 & 79.66\\
    \hline
        CAET-SWin & \textbf{82.65} [75.6 89.8] & \textbf{83.66} & 81.66\\
    \hline
    \end{tabular}
}
\end{table}
\begin{table*}[t!]
    \centering
     \caption{\footnotesize 10 fold cross-validation accuracy of $\SM$ framework and its constituent parts.    \label{tab:folds}}
    \begin{tabular}{|c|c|c|c|c|c|c|c|c|c|c|}
    \hline
        \textbf{Model} & \textbf{Fold 1} & \textbf{Fold 2} & \textbf{Fold 3}& \textbf{Fold 4}& \textbf{Fold 5}& \textbf{Fold 6}& \textbf{Fold 7}& \textbf{Fold 8}& \textbf{Fold 9}& \textbf{Fold 10}\\
    \hline
        CAET & 81.82 & 54.54 & 45.45 & 90.91 & 41.67 & 66.67 & \textbf{83.33} & 66.67 & 81.82 & 81.82\\
    \hline
        SWin & 90.91 & 72.73 & 63.64 & 100 & \textbf{91.67} & 75.00 & 58.33 & 83.33 & 72.73 & 72.73\\
    \hline
        CAET-SWin & \textbf{90.91} & \textbf{72.73} & \textbf{72.73} & \textbf{100} & 83.33 & \textbf{75.00} & 66.67 & \textbf{83.33} & \textbf{100} & \textbf{81.82}\\
    \hline
    \end{tabular}
\vspace{-.25in}
\end{table*}

\noindent
The second path of the $\SM$ framework consists of a SWin transformer (SWin-B architecture~\cite{9607582}) followed by a GMP layer that concentrates on local variations of nodules related to each subject. The Swin transformer is a hierarchical transformer that uses shifting-window MSA and includes four stages as shown in Fig.~\ref{fig:cae-transformer}. First, the input image is divided into $4\times4$ patches. In Stage 1, the feature dimension of each patch is projected into $128$ by a linear embedding layer and then SWin transformer blocks are applied on the patches. The hierarchical structure is built based on a patch merging layer at the beginning of Stages 2-4 reducing the number of patches. This layer concatenates the features of $2\times2$ neighboring patches and applies a linear layer. Afterward, several SWin transformer blocks are employed for feature transformation at each stage.

All layers of a SWin transformer are similar to the original transformer except for the MSA module, which is replaced based on a shifting window mechanism to calculate self-attention within local windows of size $\M\times\M$. The advantage of this method is to find cross-window connections and reduce computation complexity. As shown in Fig.~\ref{fig:cae-transformer}, the Swin transformer block of the $\SM$ consists of two modules, W-MSA and SW-MSA. W-MSA module is a regular partitioning scheme that divides the image into non-overlapping windows and self-attention is computed in each window. In the next step, SW-MSA computes self-attention in new windows generated by shifting the W-MSA windows by ($\lfloor \frac{M}{2} \rfloor$,$\lfloor \frac{M}{2} \rfloor$) pixels. The output of Stage 4 is the feature vector related to each slice fed to a GMP layer to aggregate the volume-level features of each subject. Here, we fine-tuned the weights of a pre-trained SWin-B transformer trained on ImageNet-21k dataset~\cite{5206848} with input image size of $224\times224$ and window size of $7\times7$.

\vspace{.025in}
\noindent
\textbf{3.3. Fusion Path}

\noindent
The output of the CAET path consists of $32$ features capturing temporal relations of CT slices associated with a  patient. On the other hand, the output of the SWin path includes $1024$ features modeling inter-slice (spatial) correlations in different CT slices of a given subject. In order to take advantage of slice information in both spatial and temporal domains, we concatenated the two output vectors to form the final feature vector, which is passed through $4$ FC layers with $512$, $128$, $32$, and $2$ neurons. The last FC layer uses a Softmax activation function to produce probability scores for the two classes.

\vspace{-.15in}
\section{Experimental Results} \label{sec:pagestyle}
\vspace{-.1in}
\setlength{\textfloatsep}{0pt}
\begin{figure}[t!]
    \centering
    \includegraphics[width=0.75\columnwidth]{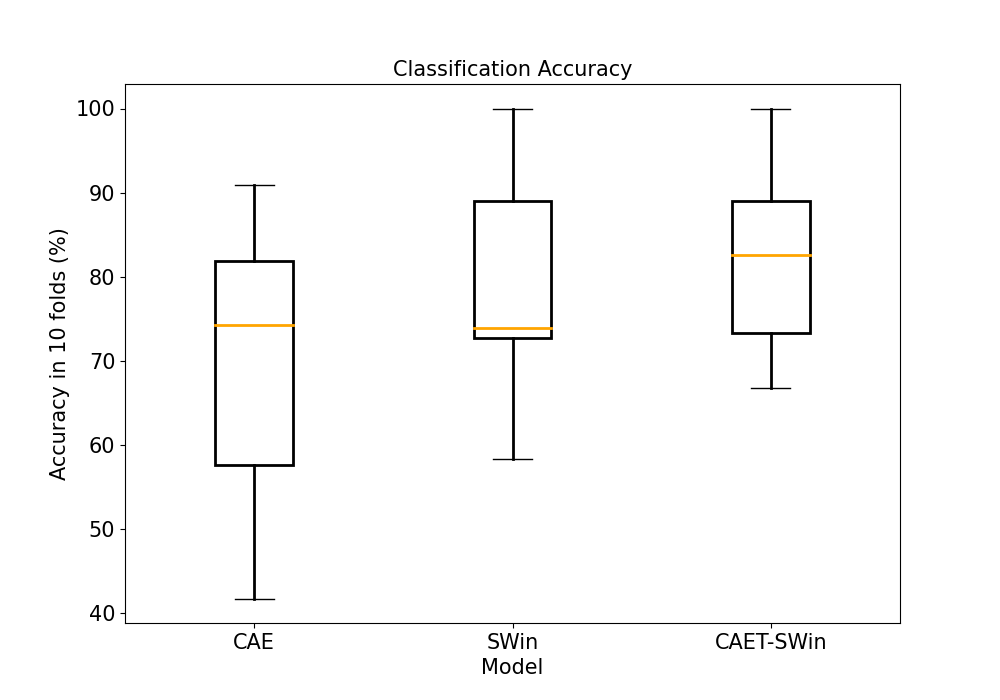}
    \vspace{-.2in}
    \caption{\footnotesize Boxplots for $\SM$ and its constituent parts.}
    \label{fig:boxplot}
\end{figure}
We evaluated the performance of the proposed $\SM$ Transformer framework using the $10$-fold cross-validation method. It should be noted that the models presented in parallel paths were trained independently. The CAE model was pre-trained using a batch size of $128$, learning rate of $1e-4$, and $200$ epochs. The best model on the randomly sampled $20\%$ of the dataset was selected as the candidate model. The model was then fine-tuned on the in-house dataset using a lower learning rate of $1e-6$ and $50$ epochs. To fine-tune the final CAE, only the middle FC layer and its previous and next convolution layers were trained while the other layers have been kept unchanged. The CAE-generated features were then used to train the transformer encoder. The transformer was trained using the Adam optimizer with a learning rate of $1e-3$, label smoothing with the $\alpha$ = $0.1$, and $200$ epochs. Simultaneously, the pre-trained SWin-B transformer was trained by AdamW optimizer with a learning rate of $1e-5$, weight-decay of $0.05$, and $50$ epochs with early stopping training strategy (patience = $10$). At the last step, FC layers were trained by means of the Adam optimizer with a learning rate of $1e-2$ and $20$ epochs. Also, dropout layers were incorporated to prevent the model from getting over-fitted. The loss function used in the whole process was cross entropy. The classification results of the $\SM$ framework are presented in Table~\ref{tab:results} in terms of accuracy, sensitivity, and specificity.

We have compared performance of the proposed $\SM$ framework with the results obtained by the model proposed in~\cite{Oikonomou2019,Gao2021} and stand-alone models in parallel paths of our hybrid model. We added a FC layer with $2$ neurons right after the last layer of each model with a Softmax activation function to classify SSNs and evaluate effects of each feature set. The experimental results provided in Table~\ref{tab:results} show that simultaneous attention to both time and space domains empowers the overall model in such a way that $\SM$ achieved the best performance among its constituent parts in all three evaluation metrics. More details regarding the performance of each fold is provided in the Table~\ref{tab:folds} and the distribution of classification accuracy for these 3 models is shown in Fig.~\ref{fig:boxplot}. Additionally, we implemented two modified versions of CAET by replacing the aggregation method with the Global Average Pooling (GAP) layer and Flattening layer. However, as presented in Table~\ref{tab:results} CAET utilizing GMP layer outperformed mentioned models. To compare the ability of the CAE algorithm in extracting efficient features fed into the transformer, we employed a pre-trained basic architecture ViT in a voting scheme presented in ~\cite{Gao2021} to classify SSNs and results demonstrated the superiority of CAET in using CAE instead of patch flattening and linear projection compared to ViT model.

As stated previously, correctly identifying the invasiveness level of nodules could have a great impact on the treatment plan and its success. Therefore, the correct diagnosis of malignant nodules is relatively more important than detection of early stages of nodule transmutation. From this point of view, sensitivity would be a more capable evaluation metric than other criteria. The $\SM$ achieved a sensitivity of $83.66\%$ which is about $4\%$ higher than the reference study~\cite{Oikonomou2019} while specificity is kept at the same value. In other words, in our hybrid model fewer cases of malignancy are missed. Hence, potentially it could be presented as a more reliable algorithm for recognizing malignant nodules. Furthermore, the confidence interval (CI) which describes the uncertainty level of a model is narrower for $\SM$ illustrating that the proposed hybrid model is more precise/reliable than its radiomics counterpart.

\vspace{-.15in}
\section{Conclusion}
\vspace{-.1in}
The paper proposed a hybrid transformer-based framework, referred to as the $\SM$, to  accurately and reliably predict the invasiveness of lung adenocarcinoma subsolid nodules from non-thin 3D CT scans. The proposed $\SM$ model achieves this objective by combining spatial (within-slice) and temporal (inter-slice) features extracted by its two constituent parallel paths (the CAET and SWin paths) designed based on the self-attention mechanism. The $\SM$ significantly improved reliability of the invasiveness prediction task compared to its radiomics-based counterpart while  increasing the accuracy by $1.65\%$ and sensitivity by $3.66\%$.  Investigating effects of embedding radiomics and morphological features in the $\SM$ framework is a fruitful direction for future research.


\bibliographystyle{IEEEbib}

\end{document}